\documentclass[11pt]{article}
\usepackage[para,online,flushleft]{threeparttable}
\usepackage{soul}
\usepackage{sidecap}
\usepackage{epsfig}
\usepackage{color}
\usepackage{graphicx}
\usepackage{multirow}
\usepackage{hanging}
\usepackage{wrapfig}
\usepackage{subfig}
\usepackage{epstopdf}
\usepackage{float}
\usepackage{booktabs}
\usepackage{algorithm}
\usepackage{algpseudocode}
\usepackage[small,compact]{titlesec}
\usepackage{palatino}
\usepackage{url}
\usepackage[rflt]{floatflt}
\usepackage{subfig}
\usepackage{comment}
\usepackage{amssymb}
\usepackage{multirow}
\usepackage{cite}
\usepackage{tabularx}
\usepackage{xcolor}
\usepackage{listings}
\usepackage{mscp}
\graphicspath{{./figures/} }
\usepackage{multicol}
\usepackage{threeparttable}
\definecolor{dkgreen}{rgb}{0,0.6,0}
\definecolor{gray}{rgb}{0.5,0.5,0.5}
\definecolor{mauve}{rgb}{0.58,0,0.82}
\newcounter{myctr}

\newcommand{\ignore}[1]{}

\usepackage{xspace}
\newcommand{\Sec}[1]{Section~\ref{#1}}
\newcommand{\Fig}[1]{Figure~\ref{#1}}

\begin{document}
\title{Dataflow Accelerator Architecture for Autonomous Machine Computing}

\author{Shaoshan Liu${}^\dagger$, Yuhao Zhu${}^\diamond$, Bo Yu${}^\dagger$, Jean-Luc Gaudiot${}^\ast$, Guang R. Gao${}^\ddagger$\\
${}^\dagger$PerceptIn Inc.\hspace{.1in} ${}^\diamond$University of Rochester\\
${}^\ast$U.C. Irvine\hspace{.1in} ${}^\ddagger$University of Delaware\\}

\date{}

\maketitle
\begin{abstract}
Commercial autonomous machines is a thriving sector, one that is likely the next ubiquitous computing platform, after Personal Computers (PC), cloud computing, and mobile computing. Nevertheless, a suitable computing substrate for autonomous machines is missing, and many companies are forced to develop \textit{ad hoc} computing solutions that are neither principled nor extensible. By analyzing the demands of autonomous machine computing, this article proposes Dataflow Accelerator Architecture (DAA), a modern instantiation of the classic dataflow principle, that matches the characteristics of autonomous machine software.
\end{abstract}

\section{Autonomous Machine: The Next Ubiquitous Computing Platform}
\label{sec:rise}

Commercial autonomous machines is a thriving sector. With a projected average compound annual growth rate (CAGR) of 26\%, by 2030 this sector will have a market size of \$1 trillion~\cite{carg}. Autonomous machine is on the verge of becoming the next ubiquitous computing platform, after personal computers, cloud computing, and mobile computing.

The continuous proliferation of autonomous machines, however, depends critically on an efficient computing substrate, driven by higher performance requirements and the miniaturization of machine form factors. Despite recent advancements in autonomous machine systems design from major industrial organizations as Google~\cite{team2019introducing}, Tesla~\cite{TeslaAutopilot}, Mobileye~\cite{Mobileye}, Nvidia~\cite{NvidiaDrive}, the computing architecture of autonomous machines still remains a largely open research question. This is because
completely independent teams have approached the problem, resulting in a bevy of solutions, some replications but certainly no consensus.

The fragmentation, while undesirable, is understandable in that
different companies target autonomous machines in different forms (e.g., cars, aerial drones, service robots, and industrial robots), each naturally differing in missions, design constraints, computational capabilities, and mechanical characteristics~\cite{liu2021rise}. 
A better understanding of the underlying issues, a formalization of the common problems, and a certain unification of the solutions would all yield a more efficient approach to the design process. 



This article first analyzes the challenges in designing an efficient computing substrate for autonomous machines, particularly focusing on why contemporary mobile Systems-on-a-Chip (SoC), a seemingly natural choice, is a bad fit. Surprisingly, classic dataflow architectures, while not enjoying practical adoption for general-purpose computing, are well-suited for autonomous machine workloads --- in principle. We propose Dataflow Accelerator Architecture (DAA), a modern instantiation of the dataflow principle in the era of hardware specialization. We describe why DAA matches the characteristics of autonomous machine workloads, and discuss key technologies that could enable DAA as a mainstream computing substrate for autonomous machines.


\section{Software Pipeline of Autonomous Machines}

\begin{figure}
    \centering
    \includegraphics[width=1.0 \textwidth]{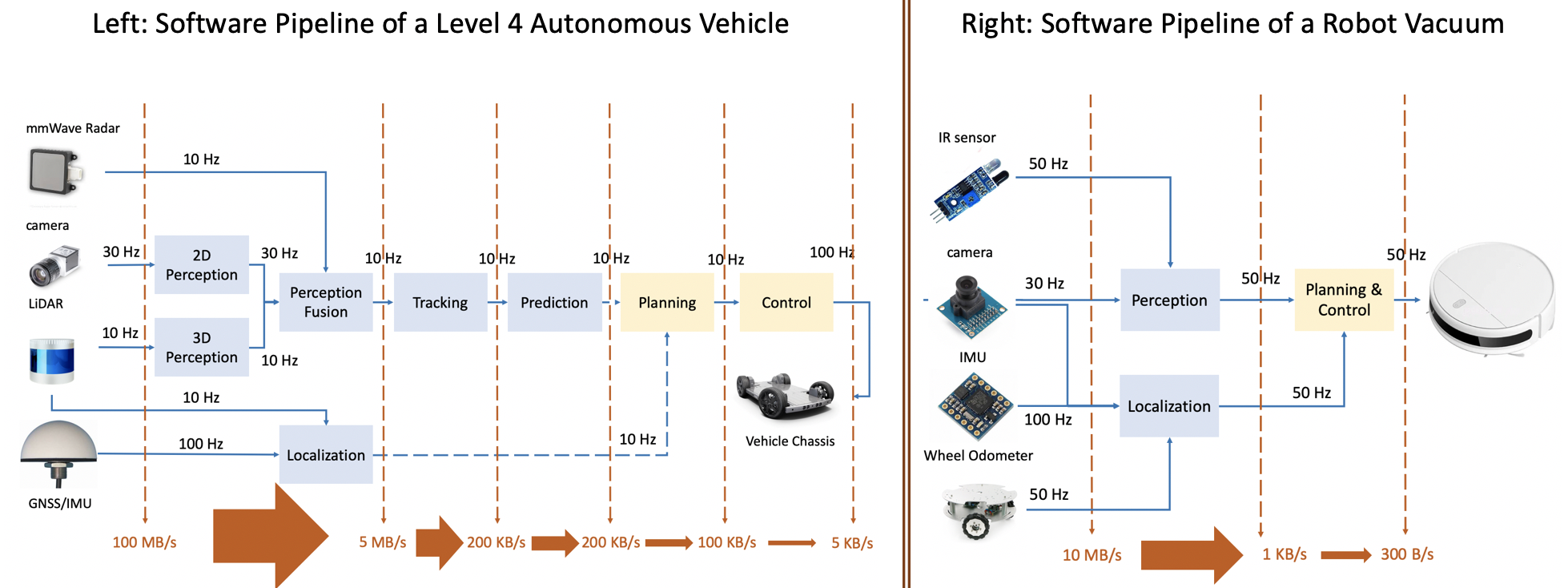}
    \caption{Left: the software pipeline of a level-4 autonomous vehicle. Right: the software pipeline of a robot vacuum. Other software organizations are possible.}
    \label{fig:am}
\end{figure}

From a thirty-thousand-foot view, the software pipeline continuously consumes data from a diverse array of sensors and produces control commands (e.g., brake, turn) to the actuator of the vehicle. \Fig{fig:am} shows the software pipelines of two representative autonomous machines, a high-end L4 self-driving car (left) and a low-end home cleaning robot (right). They differ in implementation but follow the same principle. We use an L4 autonomous vehicle we developed as a running example to describe the software pipeline.

In order to generate the control commands, the control module, the last module of the pipeline, requires a navigation plan, i.e., a path, which is generated by the path planning module. To generate a path the planning module in turn requires two pieces of information: how the environment looks like and where the agent is in the environment. The former comes from the prediction module (which predicts the motion of surrounding objects) and the latter comes from the localization module. To predict the motion of an object, the prediction module relies on the past trajectory of the object, which comes from the tracking module. The tracking module fuses the perception results from various sensors such as Radar, cameras, and LiDAR.


For the vehicle to be responsive, the software stack must generate control commands at a \textit{prescribed firing frequency} (PFF), e.g., 10 Hz in our L4 self-driving car here. This timing requirement at the output in turn translates to timing requirements at interior nodes in the M-DFG. For instance, the planning task must execute at the PFF, since planning and control are pipelined; the 2D perception task usually executes at a frequency several-fold higher than the PFF, since the tracking module uses \textit{a sequence of} 2D perception results to track objects.

\section{Why Not Use Commercial SoCs?}
\label{sec:mobile}

Commercial (mobile) SoCs would have been an ideal computing substrate for autonomous machines. The incentive is two-fold. First, since mobile SoCs have reached economies of scale, it would have been beneficial to piggyback on such affordable, backward-compatible computing systems. Second, mobile SoCs integrate domain-specific accelerators, a.k.a., Intellectual Property (IP) blocks in semiconductor parlance, which are specialized for particular algorithm domains and provide large performance and energy-efficiency benefits. In fact, mobile SoCs have been demonstrated to support mobile robots ~\cite{liu2021pi,liu2017computer}.

\begin{figure}[t]
\centering
\subfloat[Latency comparison.]
{
  \includegraphics[trim=0 0 0 0, clip, width=0.24\columnwidth]{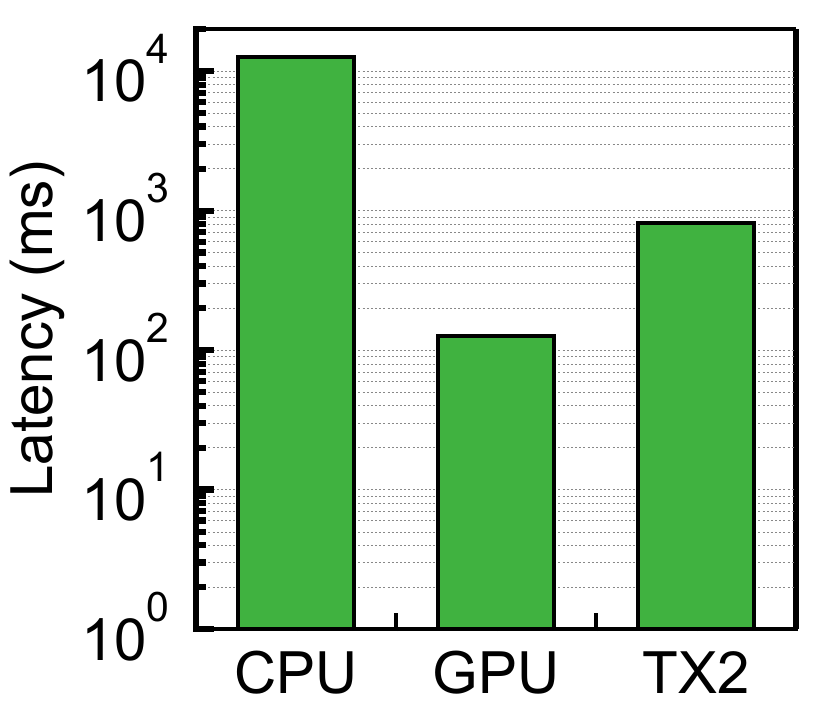}
  \label{fig:dse_l}
}
\subfloat[Energy comparison.]
{
  \includegraphics[trim=0 0 0 0, clip, width=0.24\columnwidth]{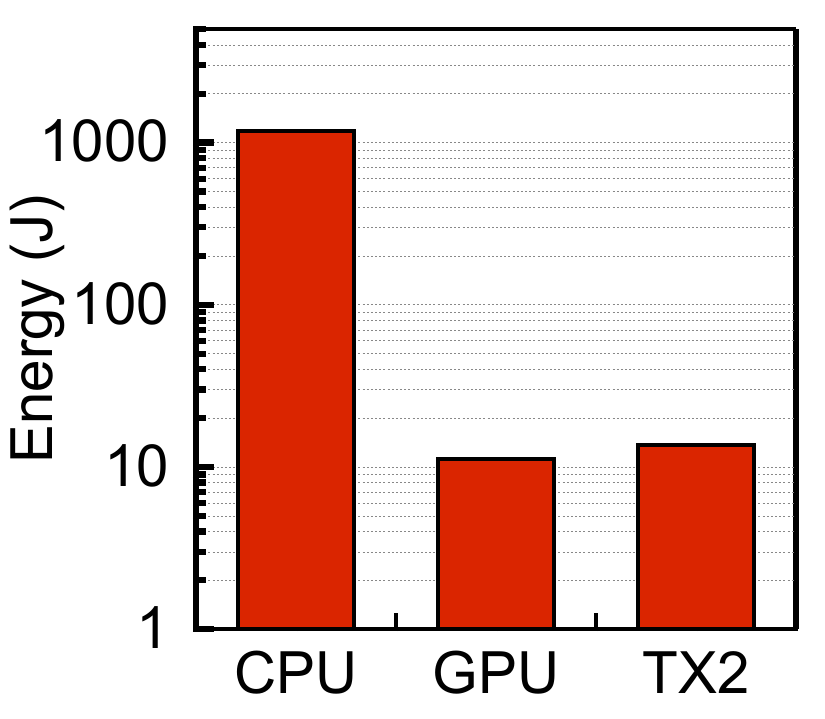}
  \label{fig:dse_e}
}
\caption{Performance and energy comparison of three platforms running the same software pipeline. On TX2, we use the Pascal GPU for depth estimation and object detection and use the ARM Cortex-A57 CPU (with SIMD capabilities) for localization. Other mappings are explored, too, but result in worse performance and/or energy-efficiency.}
\label{fig:dse}
\vspace{-10pt}
\end{figure}

We initially explored the possibility of using high-end mobile SoCs, such as the Qualcomm Snapdragon \cite{snapdragon} and Nvidia Tegra \cite{tegra}, to support autonomous driving workloads, but concluded that mobile SoCs are ill-suited for autonomous machines. \Fig{fig:dse} shows the results from one such effort using Nvidia TX2~\cite{tx2dev}, a typical mobile SoC~\cite{yu2020building}.

In this design, we map the perception task to the GPU and the rest to the CPU with extensive use of SIMD instructions. \Fig{fig:dse_l} and \Fig{fig:dse_e} show the latency and energy consumption, respectively, of an Intel Coffee Lake CPU, an Nvidia GTX 1060 GPU, and the TX2, all executing the same software pipeline. \Fig{fig:dse_l} shows that TX2 is much slower than the GPU, leading to an excessively high end-to-end latency of about 1 second. The long latency also diminishes the energy benefits of mobile SoCs. \Fig{fig:dse_e} shows that TX2 consumes more energy over the GPU. Investigating the results reveals three sources of inefficiencies.

\paragraph{Lack of Specialization.}
The processing capability of mobile SoCs is too low for realistic autonomous machine workloads due to the lack of hardware specialization. This is perhaps not all that surprising given that mobile SoCs are not built with the autonomous machine software stack in mind. Today's SoCs integrate accelerators such as GPUs, Neural Processing Units, and Digital Signal Processors (DSP), but few, if any, are dedicated to algorithms used in autonomous machines. Most of the algorithms are mapped thus to the GPU in an SoC, which serve a ``general-purpose'' accelerator. Extensive prior work shows that the speed of an autonomous machine task can be accelerated by up to one of magnitude on a dedicated accelerator over a GPU~\cite{gan2021eudoxus,suleiman2018navion}.



The lack of hardware specialization in commercial SoCs also presents a significant challenge for sensor synchronization, which is critical since autonomous machines fuse diverse sources of sensors.
Commercial SoCs lack dedicated hardware for sensor synchronization and, as a result, often synchronize sensor data in software. Software synchronization either introduces a high synchronization latency (several milliseconds) or, worse, produces incorrect synchronization results as sensor data timestamps are not directly obtained in hardware~\cite{liu2021brief,yu2020building}.


\paragraph{Inefficient Task Communication.} Mobile SoCs do not optimize the communication between hardware accelerators.
Instead, data communication requires redundant copy through the main memory, which introduces high time and power overhead for each sensor frame.
For instance, when using DSP to accelerate image processing, the CPU has to explicitly copy images from sensor interface to DSP through the entire memory hierarchy~\cite{yedlapalli2014short}.  On the Snapdragon 820 SoC, copying data through the CPU introduces a 3~ms latency overhead and a 1~W power overhead~\cite{tang2018pi}. Even worse, since many robotic systems were built on top of managed run-time environments (e.g., Android), data copying could trigger garbage collections and stall the entire pipeline~\cite{zhang2018pirvs}.

\paragraph{Centralized Task Coordination} Different accelerators are coordinated by the CPU. When a producer accelerator finishes its job, it sends an interrupt to the CPU. The CPU executes the corresponding interrupt service routine, which in turn triggers the driver of the consumer accelerator. The consumer driver, once again executes on the CPU, configures and invokes the consumer accelerator, essentially ``relaying'' the task from the producer to the consumer.

Frequently invoking the CPU prevents the CPU from entering deep sleep mode, leading to excessive energy waste. Our measurements show that the CPU services interrupts tens of times per second, whereas the CPU needs to be idle for over 1 second for it to be beneficial to enter the deep sleep mode (due to the wake-up overhead). On average, the CPU could contribute to up to half of the SoC power, offsetting the energy efficiency of accelerators.

\section{Dataflow Accelerator Architecture (DAA) to the Rescue}
\label{sec:traits}

The limitations call for a new architectural model, which we call the ``dataflow accelerator architecture'' (DAA). DAA has two ingredients. First, it incorporates a diverse set of domain specific accelerators to address the increasing performance requirement of ever more complicated algorithms. Second, DAA organizes accelerators in a dataflow fashion to remove the inefficiency of centralized coordination and communication. The dataflow organization of accelerators translates the orders of magnitude efficiency gains of individual accelerators to the end-to-end application.

This section first discusses the principles of classic data-flow architecture and how they inherently address the limitations of SoC architectures. We then discuss unique traits of the autonomous machine workloads that overcome conventional detractors that kept dataflow architectures of the past from gaining practical adoption.

\subsection{From Dataflow to Dataflow Accelerator Architectures}
\label{sec:daa}

\paragraph{Principles of Dataflow Architectures.}
Dataflow concepts originated in the 1970s and 1980s, with pioneering work by Jack Dennis, Arvind, and others~\cite{dennis1974preliminary,arvind1980dataflow}. The central idea of dataflow architectures was to do away with the classic von Neumann architecture, where instructions are executed in the explicit order specified by the control flow. Control flows limit the window in which the instruction-level parallelism (ILP) can be exploited, presenting an artificial performance roadblock. In a dataflow architecture, execution is data-driven in that an instruction, in principle, executes as soon as all its inputs are available rather than when the control flow gets to it.

Dataflow architectures rely on the dataflow graph (DFG), which captures the data dependencies and, thus, the full ILP in a program. In a DFG, each node is an instruction and edges represent direct data communications between instructions. By explicitly expressing the data dependency among instructions, a DFG allows an instruction to fire whenever its operands are ready. It is worth noting that out-of-order superscalar processors that dominate today's high-performance CPU market are essentially restricted dataflow machines that exploit ILP in a small portion (as large as the instruction window can accommodate) of the DFG~\cite{patt1985hps,palacharla1997complexity}.

\paragraph{Dataflow Accelerator Architecture.}
We see an analogy between the bottlenecks in conventional programs and those in autonomous machines software, both of which can be addressed by the dataflow principle. The key is to view the software stack of an autonomous machine, such as those in \Fig{fig:am}, as a \textit{macro dataflow graph} (M-DFG), where each node represents a high-level task such as localization and motion planning. The granularity of each task (M-DFG node) is rather coarse, usually equivalent to billions of instructions when compiled to a conventional CPU ISA.

One could view the M-DFG as an ISA for the autonomous machine software stack, and view today's SoCs as an implementation of the ISA. This implementation maps each M-DFG node to an IP block. Just like how the speed of a conventional program is limited by the control flow, the M-DFG, when executed on an SoC, is also control-limited in that the IP blocks must be centrally coordinated by the CPU and communicated through the memory, as demonstrated in \Sec{sec:mobile}.

This realization gives rise to the notion of dataflow accelerator architecture, where accelerators directly communicate with each other through dedicated \textit{on-chip buffers} and coordinate autonomously \textit{without} the CPU's intervention. This architectural model has two benefits. First, it exposes higher levels of parallelisms with each accelerator (M-DFG node) firing whenever its input data are ready. Second, it accelerates the firing rates of M-DFG nodes by making operands more readily available to consumers; this is achieved by allowing producers and consumers to directly communicate using a per-accelerator on-chip buffer rather than through the main memory.



The DAA takes inspiration from the block variant of the dataflow architecture, which partitions a dataflow graph into blocks (``megainstructions'') and either 1) executes each block in a dataflow manner and uses control flow across blocks~\cite{burger2004scaling} or 2) uses dataflow across blocks while executing each block following its control flow~\cite{sakai1989architecture}. The DAA model has a similar flavor in that each M-DFG node is essentially a block. By removing the CPU coordination and allowing M-DFG nodes to directly communicate, DAA essentially enforces the dataflow model across blocks. However, DAA does not specify how each block is executed. A block could be mapped to a CPU, executing by following the control flow, or (more commonly) mapped to a dedicated accelerator that is designed to fully exploit the ILP exposed by the dataflow.

One might find similarities between the on-chip buffers in DAA and the so-called ``system caches'' in contemporary mobile SoCs such as the Apple's A-series and Qualcomm's Snapdragon-series SoC. The main difference is that a system cache is shared across all the SoC-generated traffics without management, whereas each communication buffer in a DAA is dedicated and optimized for a pair of producer-consumer communication. Note, however, that the physical implementation of the DAA buffers can either be distributed or centralized.

\subsection{Opportunities for Autonomous Machine-Specific DAA}
\label{sec:am}

While traditional dataflow architectures target general-purpose processing, we focus on designing and optimizing DAAs specifically for autonomous machine. The software stack of autonomous machines possesses four characteristics that can be leveraged to avoid common pitfalls of traditional dataflow machines while retaining their main benefits.

\paragraph{Low Bookkeeping Overhead.} The M-DFG has only a handful of nodes, so the bookkeeping overhead, such as tag boradcasting, matching, and storage overhead, is low. In contrast, performance of conventional dataflow machines is sometimes bottlenecked by managing the tags.

\paragraph{Continuous Inputs Provide Abundant Parallelisms.} Autonomous machine workloads provide abundant parallelisms and, thus, do not typically starve the hardware. Many conventional programs do not have sufficient intrinsic parallelisms to fully utilize a realistic dataflow hardware except when processing large arrays.

The abundant parallelism comes from the fact that inputs to autonomous machines are continuous: as an autonomous machine operates, various sensors continuously pump data frames to the hardware. Since frames are largely independent, autonomous machine software exposes input-level parallelism that is unavailable in programs that conventional dataflow machines target.

\paragraph{Flexible Dependencies.}
Conversely, conventional dataflow machines are also limited when a program has \textit{too much} parallelism, which requires throttling mechanisms that degrade performance. Autonomous machine workloads, in contrast, have flexible dependencies in that a consumer node, while dependent on a producer node, can afford to drop data and operate on the \textit{latest} data from producers. As a result, parallelism explosion is general not a concern.

The reason behind the flexible dependencies in autonomous machine workloads has to do with the real-time nature of the workload. Autonomous machines must operate in real time; thus, each node in the DFG has a prescribed output frequency. For instance, the image perception node processes images at 30 FPS; the LiDAR perception node processes point clouds at 10 FPS. \Fig{fig:am} annotates each node with a firing frequency. As a result, engineers \textit{intentionally} design algorithms such that a node does not block if its firing time is reached. For instance, the planning algorithm fetches the latest produced data, essentially dropping previously produced data; the localization algorithms consumes a sequence of frames such that missing one frame of data is not catastrophic.


\paragraph{No Loops.} The M-DFG of autonomous machines is a directed acyclic graph (DAG) and the nodes within the DAG are stateless. The M-DFG nodes are thus naturally non-reentrant, eliminating the notorious difficulty to deal with reentracy in conventional dataflow architectures~\cite{dennis1983data,gurd1985manchester}.

\subsection{Programming the DAAs}
\label{sec:prog}

An architectural model is only useful when it is easy to program (and, by extension, compile for). One main detractor that kept general-purpose dataflow architectures from gaining practical adoption is that they target only functional programming languages~\cite{bird1988functional}, which naturally exposes parallelism, while providing poor support for imperative languages.


For DAA, however, programming model no longer presents a roadblock. This is because software developers for autonomous machines are already (implicitly) programming in a functional-style through the widely used Robot Operating System (ROS)~\cite{quigley2009ros}. ROS is a programming interface and run-time system for programming autonomous machines. Among many other features, ROS exposes a ``publish-subscribe'' programming interface: each task is implemented as a ROS node, which subscribes to a set of ``messages'' sent from the publishers (i.e., producers); each node is triggered whenever the messages are delivered. ROS thus explicitly encourages expressing an autonomous machine application in a data-driven manner. One could directly generate the M-DFG of a given ROS application.

\section{Timing-Safe DAAs for Autonomous Machines}
\label{sec:util}


In addition to improving the sheer performance and energy efficiency, DAA also provides a foundation for guaranteeing timing safety, i.e., meeting the (soft) real-time requirement, for autonomous machines, which challenges contemporary SoCs. This section discusses why the DAA model is a desirable substrate, followed by mechanisms that DAAs incorporate for timing safety.

\subsection{The Problem}


The real-time requirement of an autonomous machine application translates to meeting the prescribed firing frequency of each M-DFG node. Today's the computing systems provide a best-effort delivery of real-time requirement without providing any guarantees. In ROS, for instance, each task is annotated with an \textit{expected} firing frequency, which, however, is not guaranteed. We routinely see violations of the POF. For instance, in an early mobile robot deployment, while the localization frequency was expected to be 30 FPS, the achieved frame rate was on average only 20 FPS and could vary by as much as a factor of five~\cite{tang2018pi}.

\begin{figure}[t]
\centering
\subfloat[\small{Kalman gain.}]
{
  \includegraphics[trim=0 0 0 0, clip, width=0.2\columnwidth]{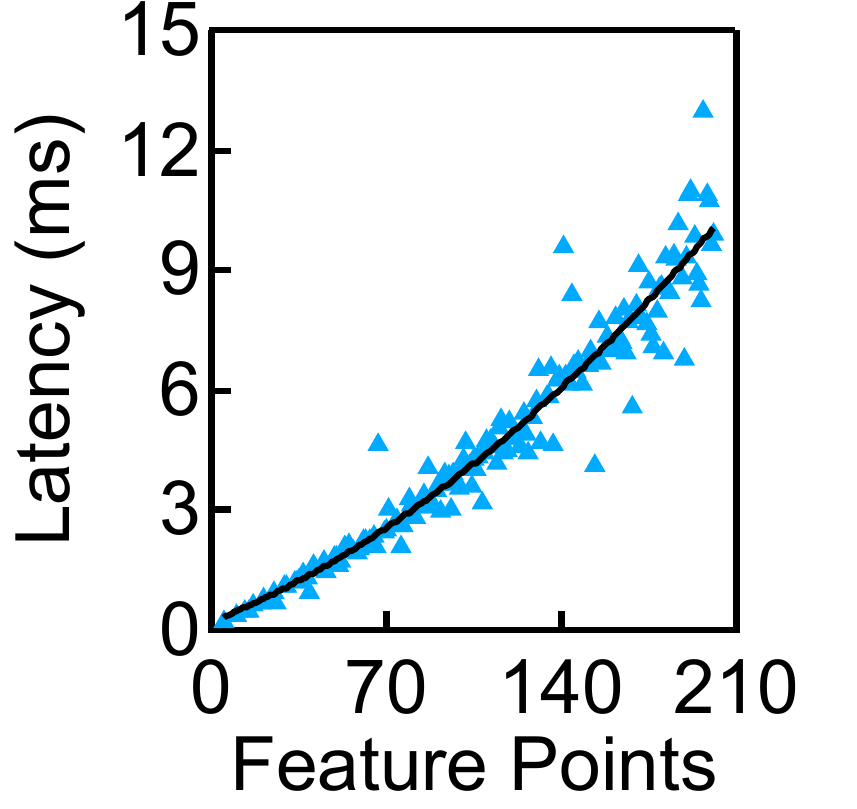}
}
\subfloat[\small{Marginalization.}]
{
  \includegraphics[trim=0 0 0 0, clip, width=0.2\columnwidth]{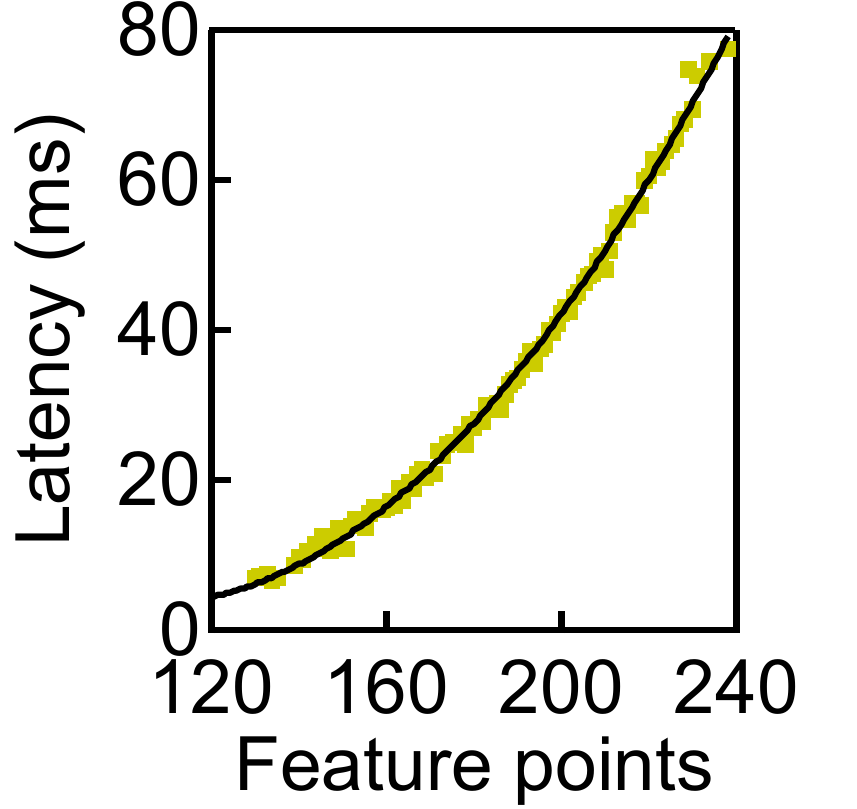}
}
\caption{Latencies of VIO's Kalman filter and marginalization blocks are correlated with the environment an agent operates in (represented by the number of visual feature points here).}
\label{fig:latency}
\end{figure}

The primary source of timing violation in SoCs is that the CPU and memory communication are known to present non-deterministic latency, susceptible to many sources of variability in the system~\cite{liu2021pi}. DAA naturally addresses this issue. DAA does not rely on CPU for task coordination and memory communication and, thus, is large free from these variabilities.

Taking the CPU and memory sideways mitigates the inter-task variabilities, but the intra-task latency, i.e., the execution latency of \textit{each task}, could be unpredictable --- for two primary reasons. First, a task's latency varies naturally with its micro-architectural implementation. Second, the environment in which a vehicle operates in changes dynamically, posing changing program inputs and computation requirements. Taking visual inertial odometery (VIO), a particular localization algorithm, as an example, Fig.~\ref{fig:latency} demonstrates that the latency of VIO's building blocks, such as Kalman filter and marginalization, varies widely but strongly correlates with the environment a vehicle operates in (represented by the number of visual feature points detected; $x$-axis)~\cite{liu2021archytas}.



\subsection{Synthesising and Dynamically Optimizing Accelerators with Timing Guarantees}

\begin{figure}[t]
  \centering
  \includegraphics[trim=0 0 0 0, clip, width=1\columnwidth]{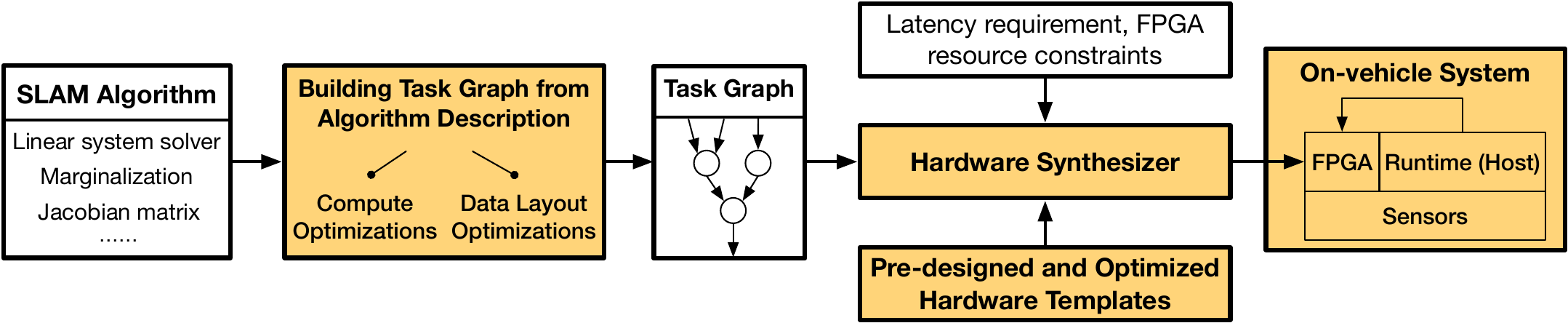}
  \caption{Archytas system overview. Archytas first generates a task graph from a high-level algorithm description. The task graph is then mapped to a parameterized hardware template, which is concretized by the hardware synthesizer (in the form of synthesizable Verilog code) given the latency and power specifications. The run-time system continuously re-optimizes the hardware according to the operating environment.}
  \label{fig:archytas}
\end{figure}

We develop a framework called Archytas to tame the intra-node variability~\cite{liu2021archytas}. Architectures work the best when the division of labor between the static and run-time systems exploits the strengths and limitations of each. Guided by this principle, Archytas integrates two components. First, a static synthesizer generates hardware accelerators to provide clean timing specifications for each task, describing clearly the per-task latency/throughput under different input conditions. Second, the run-time system reasons about the end-to-end latency using the task-level timing specification, generating timing-safe execution policies \textit{even} in dynamically changing environments. While currently targeting the localization task as a case study, we believe it generalizes to the broad autonomous machine domain. \Fig{fig:archytas} shows the Archytas framework.

\paragraph{Static Synthesis.} To reason about and guarantee timing of the entire M-DFG, we envision that each M-DFG node, when mapped to hardware, provides a latency description under different input conditions. This is achieved through a generative approach, where we synthesize an accelerator for each M-DFG node given a specific latency target (with potential power constraints). This guarantees per-task timing specification \textit{by construction}.

\begin{figure}[t]
  \centering
  \includegraphics[trim=0 0 0 0, clip, height=1.6in]{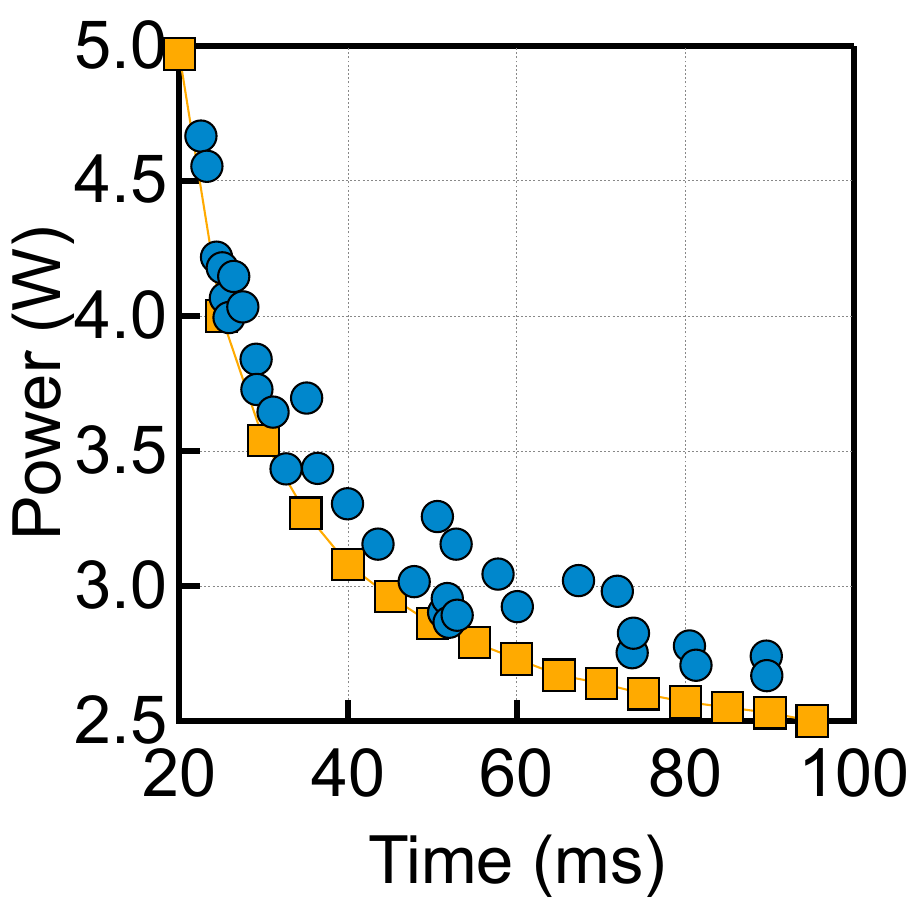}
  \caption{Latency-vs-power Pareto optimal frontier of different localization accelerators generated by Archytas.}
  \label{fig:pareto_val}
\end{figure}

Archytas analyzes an algorithm to identify a set of key architectural knobs that significantly influence the overall latency of the task. Archytas builds an analytical latency and power model of the entire algorithm parameterized by the architectural knobs.
While the latency of traditional general-purpose processors is hard to model, it is much easier to do so for accelerators due to the simpler (e.g., shallower, software-managed memory hierarchy) and more specialized design.

Using the model, hardware generation becomes a principled constraint optimization problem. A concrete, Pareto-optimal localization accelerator (in the form of synthesizable Verilog code) can be generated in just a few seconds (compared to months or even years if the design space is to be exhaustively searched) given a timing specification and power/energy constraints. \Fig{fig:pareto_val} shows the measured power-{\it vs.}-time of the Pareto-optimal accelerators generated by our constraint optimization (squares), which do indeed Pareto-dominate other designs (circles), suggesting the feasibility of generating accelerators with precise timing guarantees.


\paragraph{Dynamic Optimization.}
In dynamic environments with changing computation requirements, a purely statically-synthesized accelerator faces one of the two challenges. It either has to over-provision the hardware resources to accommodate the worst-case performance requirement, which unnecessarily wastes power in most of the time; alternatively, it could provide only an average-case guarantee without the ability to adapt to run-time dynamism.

The Archytas approach is to generate an accelerator that accommodates the average case, but couple that with a run-time system, which dynamically scales the hardware configuration to adapt to the changing workload at run time. The run-time system first detects the workload changes (e.g., using heuristics such as the number of visual features detected), and then dynamically scale up (down) the hardware capability when the workload increases (decreases). The actual mechanisms to change the hardware capability could range from light-weight clock gating to more involved techniques such as partial reconfiguration (on an FPGA platform).

\paragraph{More Advanced Roles.} In addition to the central roles described above, the compiler and run-time systems can also assume a variety of other roles, across different stages in the development and deployment cycle, when delivering timing safety. For instance, the compiler could reject timing-unsafe M-DFGs by checking the timing requirement from the autonomous machine against the timing specifications of each M-DFG node. A similar approach is a type system for constant-time cryptographic programming~\cite{cauligi2019fact}.

In addition, the compiler and run-time system can cooperate to ensure timing portability, i.e., guaranteeing timing-safety when the same code base is ported to a wide range of hardware with dramatically different computation and memory resources. The portability is critical given the enormous range of autonomous system capabilities (e.g., from home robotic vacuums all the way to autonomous cars trucks).

Consider, for example, on-chip buffer allocation and organization, which must be done in a way that does not block accelerator execution with minimal on-chip memory utilization. The compiler could exploit the observation that, under a given a set of sensors and sensing rates, the communication volume across nodes is largely deterministic. As shown in \Fig{fig:am}, the sensing module ships about 100 MB/s of data to the perception module but the output to the vehicle is only about 5 KB/s, comprised of simple actuation commands. The compiler could leverage this pattern to allocate just enough buffer space, and the run-time system could dynamically decide when and how to drop data frames to avoid timing violations (excessive stalls).

\section{Conclusion}
\label{sec:concl}

DAA offers a promising architectural model as the computing substrate for autonomous machines. Guided by the classic dataflow principle and embracing hardware specialization, DAA removes the central limitations of contemporary SoCs when executing autonomous machine workloads, and provides a desirable foundation for ensuring timing safety. DAA is also an extensible model, where new software components and hardware accelerators can be easily incorporated. The extensibility allows the DAA to potentially consolidate the fragmented hardware design space for autonomous machines.


\bibliography{refs}

\end{document}